\documentclass{aa}
\usepackage{graphicx}
\usepackage{txfonts}
\usepackage[authoryear]{natbib}
\bibpunct{(}{)}{;}{a}{}{,}

\newcommand\mcfol[1]{\multicolumn{4}{l}{#1}}

\begin{document}

\title{Multi-wavelength properties of the high-energy bright Seyfert~1 galaxy
\object{IGR~J18027$-$1455}
}


\author{J.~A. Combi\inst{1}, M. Rib\'o\inst{2,3}, J. Mart\'\i \inst{1}, S. Chaty \inst{3}}
\institute{Departamento de F\'{\i}sica (EPS), Universidad de Ja\'en,
Campus Las Lagunillas s/n, 23071 Ja\'en, Spain\\
\email{jcombi@ujaen.es}
\and Departament d'Astronomia i Meteorologia, Universitat de Barcelona, 
Mart\'{\i} i Franqu\`es 1, 08028 Barcelona, Spain
\and AIM - Astrophysique Interactions Multi-\'echelles
(UMR 7158 CEA/CNRS/Universit\'e Paris 7 Denis Diderot),
CEA Saclay, DSM/DAPNIA/Service d'Astrophysique, B\^at. 709,
L'Orme des Merisiers, 91191 Gif-sur-Yvette Cedex, France
}

\authorrunning{Combi et~al.}
\titlerunning{Multi-wavelength properties of the high-energy bright Seyfert~1 galaxy \object{IGR~J18027$-$1455}}


\date{Received / Accepted}


\abstract{A new sample of hard X-ray sources in the Galactic Plane is being
revealed by the regular observations performed by the {\it INTEGRAL}
satellite. The full characterization of these sources is mandatory to
understand the hard X-ray sky. Here we report new multifrequency radio,
infrared and optical observations of the source \object{IGR~J18027$-$1455}, as
well as a multi-wavelength study from radio to hard X-rays. The radio
counterpart of \object{IGR~J18027$-$1455} is not resolved at any observing
frequency. The radio flux density is well fitted by a simple power law with a
spectral index $\alpha=-0.75\pm0.02$. This value is typical of optically thin
non-thermal synchrotron emission originated in a jet. The NIR and optical
spectra show redshifted emission lines with $z=0.034$, and a broad H$\alpha$
line profile with FWHM $\sim$3400~km~s$^{-1}$. This suggests an Active
Galactic Nucleus (AGN) of type~1 as the optical counterpart of
\object{IGR~J18027$-$1455}. We confirm the Seyfert~1 nature of the source,
which is intrinsically bright at high energies both in absolute terms and when
scaled to a normalized 6~cm luminosity. Finally, comparing its X-ray
luminosity with isotropic indicators, we find that the source is Compton thin
and AGN dominated. This indicates that {\it INTEGRAL} might have just seen the
tip of the iceberg, and several tens of such sources should be unveiled during
the course of its lifetime.
} 

\maketitle

\section{Introduction} \label{introduction}

Unidentified high energy sources have been a subject of interest from the
early days of the {\it COS-B} era. In the 1990s, with the advent of
X-ray/$\gamma$-ray satellites like {\it ASCA} and {\it CGRO} the number of
sources with unidentified counterparts at other frequencies increased
considerably. During the first year of observations, the IBIS/ISGRI instrument
on board the {\it INTEGRAL} satellite \citep{winkler03} detected 123 hard
X-ray/$\gamma$-ray point sources, 28 of which had no clear identification with
known objects in other ranges of the electromagnetic spectrum \citep{bird04}.
These X-ray/$\gamma$-ray emitters could be high or low mass X-ray binaries,
radio quiet pulsars, clusters of galaxies, or a significant fraction of any
class of AGNs heavily obscured, at few keV, by the absorbing material of the
galactic plane. The possibility that several unidentified IBIS sources were of
extragalactic nature was early suggested by some authors (\citealt{ribo04};
\citealt{combi04}; \citealt{masetti04a}; \citealt{masetti04b};
\citealt{bassani04}; \citealt{combi05}).

\begin{figure}[t!] 
\center
\resizebox{1.0\hsize}{!}{\includegraphics[angle=0]{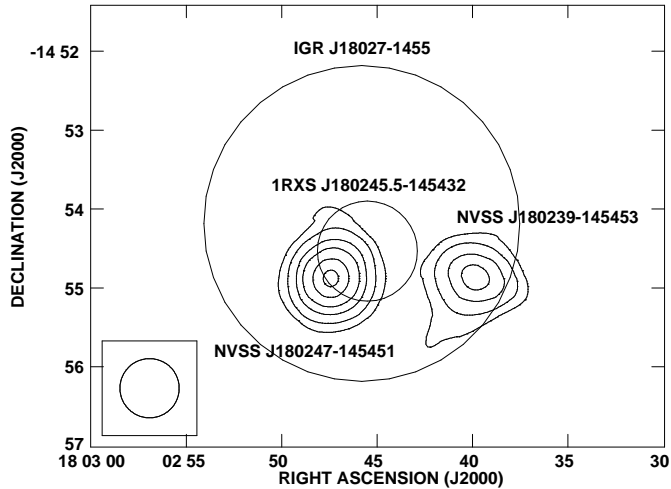}}
\caption{Image of the NVSS data obtained with the VLA at 20~cm on 1997
October 13 around \object{IGR~J18027$-$1455}. The image size is
7\farcm2$\times$5\farcm5. Contours represent $-$3, 3, 5, 8, 11, 15, 18, and 22
times the rms noise level of 0.5~mJy~beam$^{-1}$. The circle in the bottom
left corner represents the 45 arcsec of Full Width at Half Maximum (FWHM) of
the convolving beam. Two NVSS sources fall inside the 90\% error circle in
position of \object{IGR~J18027$-$1455}, and one of them is within the
2$\sigma$ uncertainty error circle of a {\it ROSAT} source.}
\label{fig:nvss}
\end{figure}

The source \object{IGR~J18027$-$1455} is one of such sources. It was 
discovered in the energy range from 20 to 100~keV during 769~ks of 
observations. Looking for possible counterparts \cite{combi04} found two 
weak point-like radio sources from the 20~cm NRAO VLA Sky Survey (NVSS, 
\citealt{condon98}) inside its 2 arcmin-radius position error circle (see 
Fig.~\ref{fig:nvss}). One of them, \object{NVSS~J180247$-$145451}, lies 
inside and near the edge of the 2$\sigma$ position error circle of the 
faint {\it ROSAT} X-ray source \object{1RXS~J180245.5$-$145432} 
\citep{voges00}, which is the only soft X-ray source well within the 
IBIS/ISGRI error circle. In addition, inside the 2$\sigma$ position error 
ellipse of this radio source, it is located an extended near infrared 
(NIR) source, \object{2MASS~J18024737$-$1454547} 
\citep{cutri03,skrutskie06}, with standard aperture magnitudes 
$J=12.78\pm0.01$, $H=11.52\pm0.01$, and $K_s=10.72\pm0.01$. Its optical 
counterpart has average magnitudes $B=19.3\pm1.0$, $R=14.9\pm0.8$ and 
$I=13.8\pm0.5$ in the USNO-B1.0 catalog \citep{monet03}. The photometry of 
the NIR/optical counterpart is not consistent with a stellar spectrum 
\citep{combi04}. On the basis of spectroscopic optical observations 
\cite{masetti04b} have tentatively classified this source as a Seyfert~1 
galaxy at redshift $z=0.035\pm0.001$.

An important characteristic of AGNs is that they radiate over a wide range of
frequencies, from radio to gamma-rays. For this reason, multi-wavelength
observations are an important tool to discriminate between objects of
different classes. Here we report multi-wavelength observations of
\object{IGR~J18027$-$1455} and discuss the obtained results. The structure of
the paper is as follows. In Sect.~\ref{observations} we describe our radio
continuum, NIR and optical observations and present the results. In
Sect.~\ref{discussion} we discuss the nature of all the detected
multi-wavelength emissions and we summarize our conclusions in
Sect.~\ref{summary}.

\section{Multi-wavelength observations and results} \label{observations}

\subsection{Radio} \label{radio}

The field of \object{IGR~J18027$-$1455} was observed with the
NRAO\footnote{The National Radio Astronomy Observatory is a facility of the
National Science Foundation operated under cooperative agreement by Associated
Universities, Inc.} Very Large Array (VLA) on 2004 December 4 and 9, with the
array being in its most extended A configuration. Each observing run lasted
for 2~h distributed among the wavelengths of 20, 6 and 3.5~cm, always with two
IF pairs of 50~MHz each. The visibilities were edited and calibrated using the
AIPS software package of NRAO. The amplitude scale was set by observing the
primary VLA calibrator \object{3C~286} and following the prescriptions
recommended in the AIPS Cookbook concerning baseline range and reduction
percentage of expected flux density as a function of the observed wavelength
and array configuration. The phase calibrator observed was the nearby source
\object{J1733$-$130} at all frequencies. The analysis of our two individual
epochs did not reveal radio sources with significant variability above the
uncertainty of our amplitude calibration. Therefore, we concatenated the
visibility data of individual epochs in order to analyze them together.
Unfortunately, the source was accidentally placed 1\farcm8 from the phase
center, so we had to correct for primary beam response using the AIPS task
PBCOR. The resulting flux densities of \object{NVSS~J180247$-$145451}, our
candidate radio counterpart to \object{IGR~J18027$-$1455}, are presented in
Table~\ref{table:vlaobs}.

\begin{table} 
\begin{center}
\caption{Observational parameters of \object{NVSS~J180247$-$145451}.}
\label{table:vlaobs} 
\begin{tabular}{c@{~~~}c@{~~~}c@{~~~}c}
\hline
\hline
$\lambda$ & Flux Density & Apparent deconvolved angular size  & P.A.\\
(cm)      & (mJy)        & ($\arcsec\times\arcsec$) & (\degr) \\
\hline
20~~      & $7.5\pm0.3$  & $(3.0\pm0.1)\times(2.1\pm0.1)$      & ~\,$95^{+5}_{-6}$\\
6         & $2.8\pm0.2$  & $(0.94\pm0.03)\times(0.27\pm0.05)$  & ~\,$90^{+2}_{-2}$\\
~~~3.5    & $2.0\pm0.1$  & $(0.56\pm0.03)\times(0.27\pm0.04)$  & $102^{+4}_{-4}$\\
\hline
\end{tabular}
\end{center}
\end{table}

\begin{figure}[t!] 
\center
\resizebox{1.0\hsize}{!}{\includegraphics[angle=0]{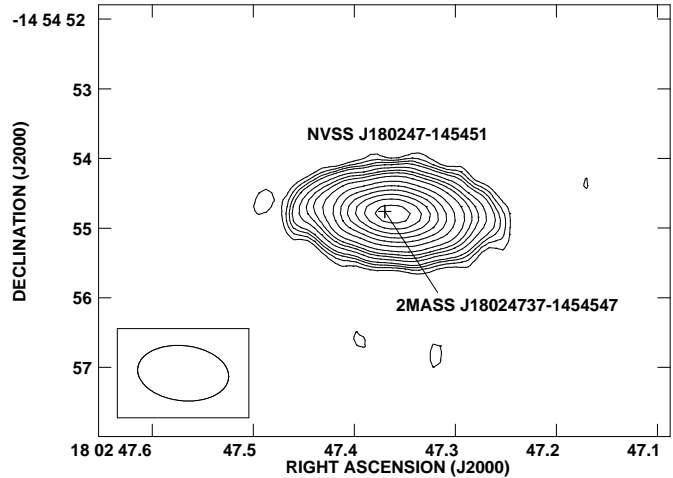}}
\caption{Naturally weighted map of NVSS~J180247$-$145451 obtained with the VLA
in A configuration on December 2004 at 6~cm. At this resolution the source has
clearly a point-like appearance. The image size is 8\arcsec$\times$6\arcsec.
The contours shown are $-$3, 3, 4, 5, 6, 8, 10, 12, 15, 20, 25, 30, 35, 40 and
45 times 32~$\mu$Jy~beam$^{-1}$, the rms noise. The ellipse shown at the
bottom left corner corresponds to a beam of 1\farcs31$\times$0\farcs79, with
position angle of 83\fdg6. The central cross marks the
location of the proposed 2MASS near infrared counterpart (3$\sigma$ uncertainty).}
\label{fig:vlamap}
\end{figure}

Our first VLA maps of \object{NVSS~J180247$-$145451} displayed a clearly
elongated radio source, as evidenced by the apparent deconvolved angular sizes
of Table~\ref{table:vlaobs}. However, this elongation is compatible with the
expected bandwidth smearing for a source located at 1\farcm8 from the phase
center \citep{taylor04}. In fact, the object was found to be consistent with
being unresolved at all frequencies. In order to avoid showing a visually
misleading figure, we present our VLA map in Fig.~\ref{fig:vlamap} with the
clean components convolved with a Gaussian beam artificially broadened in
order to approximately compensate for the expected effects of bandwidth
smearing. The corresponding beam at 6~cm wavelength is
1\farcs31$\times$0\farcs79 with position angle of 83\fdg6, which reveals the
point-like nature of the source at this resolution.

\begin{table*}[t!] 
\begin{center}
\caption{Equatorial coordinates and position uncertainties of the sources discussed in the text.}
\label{table:coord}
\begin{tabular}{lr@{$^{\rm h}$\,}r@{$^{\rm m}$\,}r@{\fs}lr@{\degr\,}r@{\arcmin\,}r@{\farcs}llll}
\hline \hline
Source    & \mcfol{$\alpha_{\rm (J2000.0~ICRS)}$} & \mcfol{$\delta_{\rm (J2000.0~ICRS)}$} & Pos. uncertainty & Astrometry \\
\hline
\object{IGR~J18027$-$1455}         & 18&02&46&0    & $-$14&54&10&0    & 2\arcmin  (90\% or 1.6$\sigma$) & 1st IBIS/ISGRI \citep{bird04} \\
\object{NVSS~J180247$-$145451}     & 18&02&47&375  & $-$14&54&54&78   & 0\farcs02 (68\% or 1$\sigma$)   & this work \\
\object{2MASS~J18024737$-$1454547} & 18&02&47&370  & $-$14&54&54&76   & 0\farcs03 (68\% or 1$\sigma$)   & 2MASS \citep{cutri03} \\
\object{USNO-B1.0~0750-0506536}    & 18&02&47&38   & $-$14&54&55&1    &  0\farcs2  (68\% or 1$\sigma$)  & USNO-B1.0 \citep{monet03} \\
\object{1RXS~J180245.5$-$145432}   & 18&02&45&5    & $-$14&54&32&0    & 19\farcs0 (68\% or 1$\sigma$)   & {\it ROSAT} \citep{voges00} \\
\hline
\end{tabular}
\end{center}
\end{table*}

Our best estimate of the radio source position is $\alpha_{\rm
J2000.0}=18^{\rm h} 02^{\rm m} 47\fs375\pm0\fs003$ and $\delta_{\rm
J2000.0}=-14\degr 54\arcmin 54\farcs78\pm0\farcs02$, determined from 3.5~cm
observations. The positions of the objects at different wavelengths are
summarized in Table~\ref{table:coord}. The radio spectrum is well fitted by a
simple power law $S_{\nu} = (9.7 \pm 0.3~{\rm mJy}) (\nu/{\rm GHz})^{-0.75 \pm
0.02}$.

\subsection{Near to far infrared} \label{infrared}

We conducted photometric ($J$, $H$ and $K_s$ filters) and spectroscopic
(0.9--2.5~$\mu$m) NIR observations of \object{2MASS~J18024737$-$1454547} on
2004 July 10 with the spectro-imager SofI, installed on the ESO New Technology
Telescope (NTT). We used the large field imaging of SofI's detector, giving an
image scale of 0\farcs288 pixel$^{-1}$ and a field of view of
4\farcm94$\times$4\farcm94. Concerning the photometric observations, we
repeated a set of observations for each filter with 9 different 30\arcsec\
offset positions including \object{2MASS~J18024737$-$1454547}, with an
integration time of 90 seconds for each exposure, following the standard
jitter procedure allowing to cleanly subtract the blank sky emission in NIR.
We observed two photometric standard stars of the faint NIR standard star
catalog of \cite{persson98}: sj9157 and sj9172. 

\begin{figure}[t!] 
\center
\resizebox{1.0\hsize}{!}{\includegraphics[angle=0]{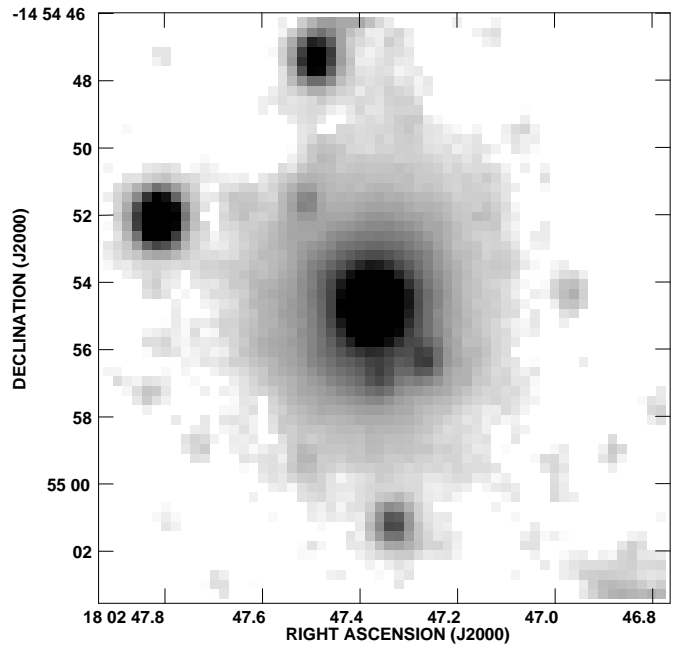}}
\caption{$K_s$ band image of the region around
\object{2MASS~J18024737$-$1454547} obtained with the NTT on 2004 July 10. A
square root transfer function has been used to enhance the faint emission. The
extended nature of the NIR source, the host galaxy, is clearly visible, mostly
aligned in the north-south direction. The image size is
17\arcsec$\times$18\arcsec.}
\label{fig:ks}
\end{figure}

We used the IRAF (Image Reduction and Analysis Facility package) suite to
perform the data reduction, including flat-fielding and NIR sky subtraction.
For the three obtained images, one in each filter, we obtained an astrometric
solution by using more than 200 coincident 2MASS objects, with a final rms of
0\farcs07 in each coordinate. We show the final $K_s$ band image in
Fig.~\ref{fig:ks}, where the extended nature of
\object{2MASS~J18024737$-$1454547} can be easily seen. We carried out aperture
photometry and transformed the instrumental magnitudes into apparent
magnitudes with the standard relation: ${\rm mag_{app} = mag_{inst} - Zp
-~}{ext\times AM}$ where $\rm mag_{app}$ and $\rm mag_{inst}$ are respectively
the apparent and instrumental magnitudes, Zp is the zero-point, $ext$ the
extinction and $AM$ the airmass. The observations were performed through an
airmass close to 1. We obtained two sets of measurements for different
appertures to include only the nucleus or the nucleus plus all the extended
emission. Using a 4-pixel aperture diameter (1\farcs2), and an adjacent
annulus with outer radius of 5 pixels to estimate the sky background, we
obtain: $J=14.9\pm0.1$, $H=13.3\pm0.1$ and $K_s=11.7\pm0.1$. The magnitudes of
the whole extended emission (nucleus+host galaxy) are: $J=12.77\pm0.02$,
$H=11.41\pm0.03$ and $K_s=10.44\pm0.04$. These last values are compatible
within errors to those present in the 2MASS catalog, obtained with a 4\arcsec\
aperture, except for the $K_s$ band, where we find a slightly brighter source.

\begin{figure}[t!] 
\center
\resizebox{1.0\hsize}{!}{\includegraphics[angle=0]{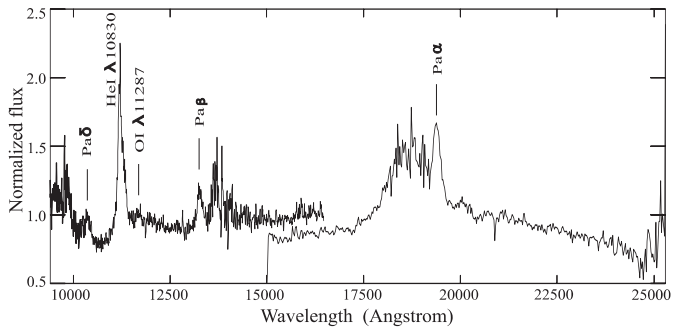}}
\caption{Near infrared spectrum of \object{2MASS~J18024737$-$1454547}, the
counterpart of \object{IGR~J18027$-$1455}, acquired with the NTT on 2004 July
10. The identified emission lines are indicated. A redshift of
$z=0.034\pm0.001$ is obtained. The strong noise redwards of Pa$\beta$ and
bluewards of Pa$\alpha$ is due to telluric features.}
\label{fig:nir_spec}
\end{figure}

Concerning the NIR spectroscopy, we took 12 spectra with the Blue and Red 
grisms, respectively. The position of \object{2MASS~J18024737$-$1454547} 
in the slit was offset 30\arcsec\ in half of the exposures to subtract the 
blank NIR sky. The total integration time was 240~s in each grism. We took 
Xe lamp exposures to perform the wavelength calibration. We extracted the 
spectra using the IRAF {\sc noao twodspec} package. 
Figure~\ref{fig:nir_spec} shows the obtained normalized spectrum with the 
most important NIR emission lines indicated. We were unable to completely 
remove the telluric features between 1.8 and 1.93~$\mu$m, due to the 
atmospheric absorption. Several lines as Pa$\delta$, \ion{He}{i} 
$\lambda$10830, \ion{O}{i} $\lambda$11287 (clearly detected after binning 
the spectrum), Pa$\beta$ and Pa$\alpha$ are clearly visible. Using these 
NIR emission lines we obtain a redshift to the source of 
$z=0.034\pm0.001$. We found no Br$\gamma$ emission nor any [\ion{Fe}{ii}] 
component.

We finally cross checked the position of the radio source 
NVSS~J180247$-$145451 with the {\it 
IRAS}\footnote{http://irsa.ipac.caltech.edu/applications/Gator/} Point 
Source catalog. At the mid-far infrared part of the spectrum, from 12 to 
100 microns, we found no counterpart to our target. A total of 1168 {\it 
IRAS} sources lie in a circle of 3\degr\ radius centered around the radio 
position. As explained in \cite{filliatre04}, we estimate as a robust 
upper limit for our target the flux such as 90\% of the {\it IRAS} sources 
have greater fluxes, leading to 0.40, 0.35, 0.72, 9.1~Jy at 12, 25, 60 and 
100 microns, respectively.

\subsection{Optical} \label{optical}

\begin{figure}[t!] 
\center
\resizebox{1.0\hsize}{!}{\includegraphics[angle=0]{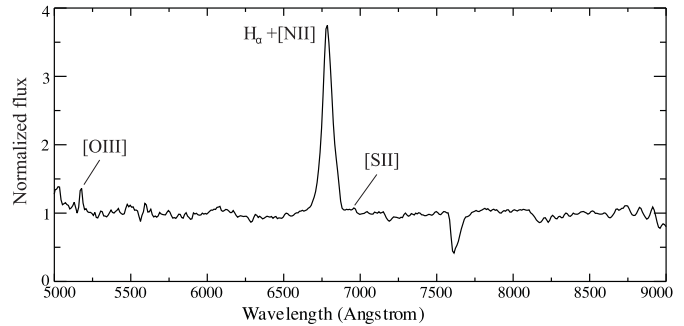}}
\caption{Averaged optical spectrum of the optical counterpart of
\object{IGR~J18027$-$1455} acquired with CAFOS on the 2.2m telescope at CAHA
on 2004 July 19. The spectrum has been smoothed with a Gaussian filter. The
identified emission lines are indicated. A broad H$\alpha$ line dominates the
spectrum. A redshift of $z=0.034\pm0.001$ is obtained.}
\label{fig:caha}
\end{figure}

The optical counterpart candidate of \object{IGR~J18027$-$1455}, namely 
\object{USNO-B1.0~0750-0506536}, was observed in 2004 July 19 with the 
2.2~m telescope of the Centro Astron\'omico Hispano Alem\'an (CAHA) in 
Calar Alto (Spain) under Director Discretionary Time (DDT). We used the 
CAFOS spectrograph with the grism R400 initially selected based on 
sensitivity criteria. This instrumental setup provided a dispersion of 
9.65~\AA~pixel$^{-1}$ with the CCD detector being a SITE\#1d\_15 chip. 
Three science exposures, of 20~min each, were taken and combined into a 
single spectrum. The frames were bias subtracted, flatfielded and 
wavelength calibrated using HgCd/He/Rb lamps and the IRAF software 
package. No flux calibration was obtained during our DDT observations and, 
therefore, the resulting spectrum in Fig.~\ref{fig:caha} is presented in 
normalized units. The optical spectrum is strongly dominated by H$\alpha$ 
emission with $z=0.034\pm0.001$ and a broad profile with a 
FWHM~$\simeq3400\pm300$~km~s$^{-1}$ (mean between Gaussian and Lorentzian 
fits without deblending of the [\ion{N}{ii}] lines).

\section{Discussion} \label{discussion}

The position agreement between the \object{IGR~J18027$-$1455},
\object{1RXS~J180245.5$-$145432} and \object{NVSS~J180247$-$145451} indicates
that these are the multi-wavelength manifestations of the same source
radiating at different bands of the electromagnetic spectrum. Moreover, the
recently published second IBIS/ISGRI catalog \citep{bird06} lists a position
and error circle for \object{IGR~J18027$-$1455} that clearly exclude the other
radio source, namely \object{NVSS~J180239$-$145453}, as a possible counterpart
(even at the 90\% confidence level). The NIR source
\object{2MASS~J18024737$-$1454547} has a position in agreement within errors
with our new precise radio position of \object{NVSS~J180247$-$145451}. Its
optical counterpart is \object{USNO-B1.0~0750-0506536}. Both the NIR and
optical objects are clearly extended (this work and \citealt{masetti04b},
respectively). In addition, the redshifted emission lines seen in the NIR and
optical spectra reveal an extragalactic object at a redshift of
$z=0.034\pm0.001$, with a broad H$\alpha$ emission line with
FWHM~$\simeq3400\pm300$~km~s$^{-1}$ (these values are compatible with those of
\citealt{masetti04b}, although they reported a slightly narrower emission line
with FWHM~$\sim$2700~km~s$^{-1}$). We are therefore seeing the broad line
region (BLR), and the object is classified as a type~1 AGN.

The steep radio spectral index ($\alpha=-075\pm0.02$) strongly supports a
non-thermal emission mechanism of synchrotron nature. This is clearly
compatible with optically thin extended jet emission from an extragalactic
source. On the other hand, the NVSS flux density of the source at 1.4~GHz is 
$10.5\pm0.6$~mJy, to be compared with our measurement of $7.5\pm0.3$~mJy. Both
values are only marginally consistent at the 3$\sigma$ level, suggesting that
the source is variable at radio wavelengths. We note that 0.61~GHz (49~cm
wavelength) observations conducted 3 months later provided a detection at a
level of $5.0\pm0.35$~mJy \citep{pandey06}, either supporting the variability
of the source or indicating that there is a low frequency turnover.

The NIR spectrum is very similar to other well studied Seyfert~1 galaxies 
such as \object{NGC~863} or \object{Mrk~335} \citep{rodriguez02}, although 
the poor signal-to-noise ratio of our NIR observations is not enough to 
discriminate weak lines as in these cases. It is interesting to note that 
the permitted \ion{O}{i} $\lambda$11287 line is a feature completely 
associated with the BLR of Seyfert galaxies. In our case this line is 
marginally detected, as in \object{NGC~863} \citep{rodriguez02}. The 
non-detection of the Br$\gamma$ line seems to suggest that thermal UV 
heating is not important, as it also happens in the case of 
\object{NGC~1097} \citep{reunanen02}.

The multiwavelength properties of \object{IGR~J18027$-$1455} strongly 
support an AGN nature and more specifically a type~1 Seyfert galaxy. In 
order to compare the broadband emission of the object with that one of the 
mean for type~1 Seyfert galaxies, we have determined the nuclear spectral 
energy distribution (SED), from the radio to the gamma-ray band. The 
observations used to build the SED have been discussed in 
Sect.~\ref{observations}. The observed magnitudes in the NIR and optical 
bands were corrected for reddening from Galactic extinction based on the 
estimated hydrogen column density $N_{\rm H}=(5.0\pm1.0) \times 10^{21}$ 
cm$^{-2}$ \citep{dickey90} and the \cite{predehl95} relationship $A_V = 
(5.59\pm0.10) \times 10^{-22}~N_{\rm H}$, which gives $A_V=2.8\pm0.6$ 
magnitudes. The transformation between the absorption in the optical $A_V$ 
and that at other wavelengths was computed according to the \cite{rieke85} 
interstellar extinction law.

At soft X-ray energies, between 0.1--2.4~keV, the flux was obtained using 
the {\it ROSAT}/PSPC count rate of 
$(3.25\pm1.39)\times10^{-2}$~count~s$^{-1}$ \citep{voges00} and a photon 
index of $\Gamma=+1.9\pm01$, typical of Seyfert~1 galaxies 
\citep{malizia03}. We used the web based tool PIMMS~v3.7a\footnote{\tt 
http://heasarc.gsfc.nasa.gov/Tools/w3pimms.html} and the $N_{\rm H}$ value 
given above to obtain an unabsorbed flux of 
$2.4^{+2.0}_{-1.3}\times10^{-12}$~erg~cm$^2$~s$^{-1}$ (propagating all 
possible uncertainties). In addition, extrapolation of the {\it 
ROSAT}/PSPC count rate with the same photon index and absorption, 
considering all possible uncertainties, provides a flux of 
$1.5^{+1.4}_{-0.9}\times10^{-12}$ erg~cm$^{-2}$~s$^{-1}$ in the 2--10~keV 
energy range.

The average fluxes detected by {\it INTEGRAL} in the 20--40 and 40--100~keV
energy ranges are $3.0\pm0.2$ and $3.3\pm0.3$~mCrab, respectively
\citep{bird06}. These can be converted to the cgs fluxes (assuming a Crab-like
spectrum and the values in \citealt{bird06})
$(2.3\pm0.2)\times10^{-11}$~erg~cm$^2$~s$^{-1}$ and
$(3.1\pm0.3)\times10^{-11}$~erg~cm$^2$~s$^{-1}$, respectively, which provide a
total flux in the 20--100~keV range of
$(5.4\pm0.4)\times10^{-11}$~erg~cm$^2$~s$^{-1}$. We note that an analysis with
more {\it INTEGRAL} data reveals the following values in the 20--100~keV range
\citep{bassani06}: $2.6\pm0.1$~mCrab and
$(4.4\pm0.2)\times10^{-11}$~erg~cm$^2$~s$^{-1}$. As can be seen there are
hints of variability in the hard X-ray/gamma-ray flux of
\object{IGR~J18027$-$1455}, and the average of
$(4.9\pm0.5)\times10^{-11}$~erg~cm$^2$~s$^{-1}$ will be used when plotting the
SED.

To compute the monochromatic luminosities we have adopted the cosmological 
parameters from \cite{spergel03}: $H_{0}=71$~km~s$^{-1}$ Mpc$^{-1}$, 
$\Omega_{\Lambda}=0.73$ and $\Omega_{\rm m}=0.27$. Using our measured 
redshift of $z=0.034\pm0.001$ we obtain\footnote{\tt 
http://www.astro.ucla.edu/$\sim$wright/CosmoCalc.html} a luminosity 
distance of $147\pm5$~Mpc for \object{IGR~J18027$-$1455}, leading to a 
hard X-ray luminosity of $(1.3\pm0.1)\times10^{44}$~erg~s$^{-1}$. The 
source is one of brightest Seyfert~1 galaxies detected so far: it is 
brighter than any of the Seyfert~1 galaxies detected with {\it BeppoSAX} 
\citep{panessa04t}, and the 4th brightest one among the 14 detected with 
{\it INTEGRAL} \citep{bassani06}.

\begin{figure}[t!] 
\center
\resizebox{1.0\hsize}{!}{\includegraphics[angle=0]{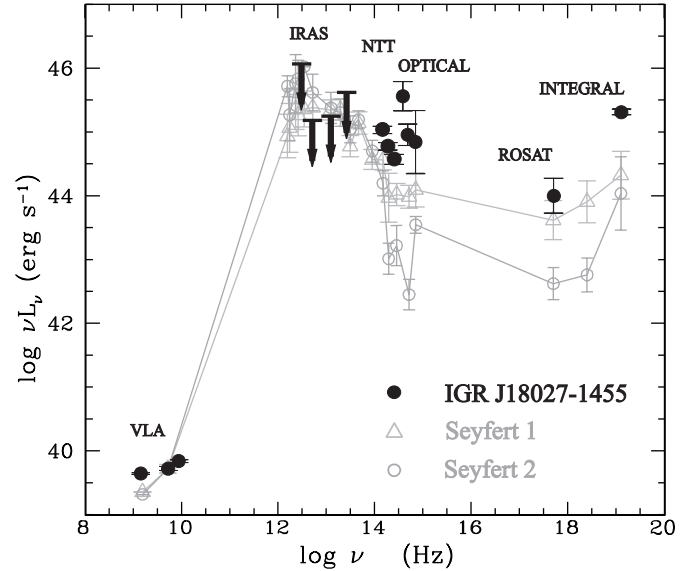}}
\caption{Overall nuclear spectral energy distribution of 
\object{IGR~J18027$-$1455} (filled symbols) from the radio to the hard 
X-ray/gamma-ray band. Optical stands for the $I$ and $B$ USNO-B1.0 and $R$ 
magnitude from \cite{masetti04b}. The SED has been normalized at 6~cm 
(assuming $\log(\nu L_{\nu}[{\rm erg~s}^{-1}])=39.75$) in order to compare 
between type 1 and type~2 Seyfert galaxies (open symbols; adapted from 
\citealt{panessa04t}; vertical bars are errors of the averages, not the 
standard deviation of the samples). The real luminosities of our target 
source are 1.21 dex lower than those shown. The SED of 
\object{IGR~J18027$-$1455} resembles the average one for Seyfert~1 
galaxies, although it is brighter in the hard X-ray domain. Optical data 
is probably contaminated by the host galaxy.}
\label{fig:sed}
\end{figure}

We show in Fig.~\ref{fig:sed} the overall nuclear SED of 
\object{IGR~J18027$-$1455} in a $\log(\nu)$--$\log(\nu L_{\nu})$ 
representation, but normalized at 6~cm for comparison with the mean SEDs 
of Seyfert galaxies \citep{panessa04t,panessa04}. The real luminosities of 
\object{IGR~J18027$-$1455} are 1.21 dex lower than those shown. Although 
in the NIR domain we have used nuclear magnitudes, the optical magnitudes 
($I$ and $B$ from USNO-B1.0 and $R=16.55\pm0.01$ from 
\citealt{masetti04b}) could be strongly contaminated by the host galaxy 
due to a limited angular resolution of $\sim$1\arcsec~pixel~$^{-1}$. We 
have not plotted the extrapolated 2--10~keV flux. Although the archival 
{\it IRAS} data does not allow us to detect the far infrared bump, the 
NIR/{\it ROSAT} data clearly show a SED typical of Seyfert~1 galaxies. 
However, the {\it INTEGRAL} data show that the source is clearly 
high-energy bright. Therefore, \object{IGR~J18027$-$1455} is not only 
among the brightest type~1 Seyfert galaxies at high energies in absolute 
terms, but also when normalized to the 6~cm luminosity.

In the basic scheme for unification of AGNs, Seyfert galaxies are divided in
two class. Those that have narrow forbidden lines and a BLR in their optical
spectrum (Seyfert~1) and those that only have narrow lines (Seyfert~2). While
the broad lines originate near the central massive black hole located at
$\leq0.1$~pc, narrow lines arise far from the nuclear engine at a distance
$\leq100$~pc. Specifically they are the same type of object but, according to
the standard model, in Seyfert~2 galaxies the BLR is obscured by a molecular
torus \citep{antonucci93}. For this reason, the majority of these objects are
Compton thick, that is, the medium is thick to Compton scattering so that the
transmitted component is dramatically suppressed below 10~keV down to the NIR
domain.

We can thus further check the Seyfert~1 nature of 
\object{IGR~J18027$-$1455} and its agreement with unification schemes by 
comparing its soft X-ray luminosity with isotropic indicators. This allows 
us to discriminate if starburst or AGN is the dominant component, and at 
the same time to assess if the source is Compton thin or Compton thick. If 
the presence of a molecular torus around the central region is important, 
the X-ray emission coming from the central engine will be negligible and 
it should be coming from a more extended zone like the NLR or a starburst 
region. In this case, the column density could be obtained indirectly from 
the flux ratio between the X-ray flux and isotropic emission measurements 
like the [\ion{O}{iii}]$\lambda$5007 and far-infrared fluxes. Both are 
good isotropic indicators, while [\ion{O}{iii}]$\lambda$5007 emission is 
produced by photons originated in the central nucleus, infrared emission 
is mainly associated to star-forming activity, and therefore produced in a 
larger region than that of the molecular torus.

To compute the [\ion{O}{iii}]$\lambda$5007 flux we have used our 
normalized optical spectrum and the average optical spectrum of 
\cite{masetti04b}. The line is clearly detected at $\lambda$=5175~\AA. 
Smoothing our spectrum with a Gaussian function, its equivalent width is 
$10\pm2$~\AA\ and the [\ion{O}{iii}]$\lambda$5007 flux is 7.1 $\times$ 
10$^{-16}$ erg~cm$^{-2}$~s$^{-1}$. This flux has been corrected for 
extinction using the equation given by \cite{bassani99}. Using a H$\alpha 
\cong$ 8.5 $\times$ 10$^{-16}$ erg~cm$^{-2}$~s$^{-1}$~\AA$^{-1}$ and a 
H$\beta \leq$ 0.5 $\times$ 10$^{-16}$ erg~cm$^{-2}$~s$^{-1}$~\AA$^{-1}$, 
the observed flux ratio H$\alpha$/H$\beta$ $\geq$ 17 and $F_{\rm 
[OIII],~cor} \geq$ 1.2 $\times$ 10$^{-13}$ erg~cm$^{-2}$~s$^{-1}$. Since 
the $F_{\rm X}$, between 2--10~keV, is 0.6--2.9 $\times$ 10$^{-12}$ 
erg~cm$^{-2}$~s$^{-1}$ the $F_{\rm X}/F_{\rm [OIII]}$ ratio is between 5 
and 24. According to the flux diagnostic diagrams introduced by 
\cite{panessa04t} for type 1 and 2 Seyferts, the source is Compton thin. 
To calculate the far-infrared flux we have adopted the equation of 
\cite{mulchaey94}. As a result the infrared flux is $F_{\rm IR} \leq$ 8 
$\times$ 10$^{-11}$ erg~cm$^{-2}$~s$^{-1}$. Therefore, the flux ratio 
$F_{\rm [OIII]}/F_{\rm IR} \geq$ 1.5 $\times$ 10$^{-3}$. According to 
\cite{panessa04t}, this shows that AGN contribution, not starburst, is the 
dominant component.

\section{Summary} \label{summary}

We can summarize our main results as follows:

\begin{enumerate}

\item The radio counterpart of \object{IGR~J18027$-$1455} has not been 
resolved at any frequency by the VLA in its most extended A configuration. 
The radio flux density is well fitted by a simple power law with a 
spectral index $\alpha=-0.75\pm0.02$, typical of optically thin 
synchrotron radiation originated by a non-thermal jet. VLBI observations 
are needed to resolve the expected jet-like structure of this high energy 
object.

\item The NIR spectrum of the \object{IGR~J18027$-$1455} counterpart shows 
several emission lines with a redshift $z=0.034\pm0.001$. The optical 
spectrum is strongly dominated by redshifted H$\alpha$ emission with 
$z=0.034\pm0.001$ and a broad profile with 
FWHM~$\simeq3400\pm300$~km~s$^{-1}$. This confirms that 
\object{IGR~J18027$-$1455} is a type~1 AGN.

\item We determined the nuclear SED of \object{IGR~J18027$-$1455} using new
radio, optical and infrared observations and available data in the literature.
Comparing our obtained SED with the average ones of Seyfert galaxies, we found
that it is typical of Seyfert~1 galaxies, although brighter than the mean at
high energies, both in absolute terms and when using a normalized 6~cm
luminosity.

\item We checked independently the Seyfert~1 nature of
\object{IGR~J18027$-$1455} by comparing its X-ray luminosity with isotropic
indicators. The source is Compton thin, as expected in unification schemes,
and the AGN contribution is the dominant component.

\end{enumerate}

\begin{acknowledgements}

J.A.C. is a researcher of the programme {\em Ram\'on y Cajal} funded jointly
by the Spanish Ministerio de Educaci\'on y Ciencia (MEC) and Universidad 
de Ja\'en.
M.R. has been supported by the French Space Agency (CNES)
and by a Marie Curie Fellowship of the European Community programme
Improving Human Potential under contract number HPMF-CT-2002-02053, and is 
being supported by a {\em Juan de la Cierva} fellowship from MEC.
The authors also acknowledge support by DGI of MEC under grants 
AYA2004-07171-C02-02 and AYA2004-07171-C02-01, FEDER funds and Plan 
Andaluz de Investigaci\'on of Junta de Andaluc\'{\i}a as research group 
FQM322.
Partly based on observations collected at the European Southern 
Observatory, Chile (observing proposal ESO N$^{\rm o}$ 073.D-0339), and on 
observations collected at the Centro Astron\'omico Hispano Alem\'an (CAHA)
at Calar Alto, Spain, operated jointly by the Max-Planck Institut f\"ur
Astronomie and the Instituto de Astrof\'{\i}sica de Andaluc\'{\i}a (CSIC).


\end{acknowledgements}

\end{document}